\documentclass{webofc}
\usepackage[varg]{txfonts}   
\usepackage{psfrag}
\usepackage{graphics}
\usepackage{float}
\usepackage{amssymb,epsfig,amsmath,euscript,array}
\usepackage{multicol}
\usepackage{subfig}
\usepackage{fancyhdr}
\usepackage{color}
\usepackage{scalerel,stackengine}

\newcommand{\be}{\begin{equation}}
\newcommand{\ee}{\end{equation}}

\newcommand{\bm}[1]{\mbox{\boldmath $#1$}}

\def\bd{\begin{document}}
\def\ed{\end{document}}
\def\nn{\nonumber}
\def\bea{\begin{eqnarray}}
\def\eea{\end{eqnarray}}
\let\bm=\bibitem
\let\la=\label

\def\npb#1#2#3{Nucl. Phys. {\bf{B#1}} #3 (#2)}
\def\plb#1#2#3{Phys. Lett. {\bf{#1B}} #3 (#2)}
\def\prl#1#2#3{Phys. Rev. Lett. {\bf{#1}} #3 (#2)}
\def\prd#1#2#3{Phys. Rev. {D \bf{#1}} #3 (#2)}
\def\cmp#1#2#3{Comm. Math. Phys. {\bf{#1}} #3 (#2)}
\def\cqg#1#2#3{Class. Quantum Grav. {\bf{#1}} #3 (#2)}
\def\nppsa#1#2#3{Nucl. Phys. B (Proc. Suppl.) {\bf{#1A}}#3 (#2)}
\def\ap#1#2#3{Ann. of Phys. {\bf{#1}} #3 (#2)}
\def\ijmp#1#2#3{Int. J. Mod. Phys. {\bf{A#1}} #3 (#2)}
\def\rmp#1#2#3{Rev. Mod. Phys. {\bf{#1}} #3 (#2)}
\def\mpla#1#2#3{Mod. Phys. Lett. {\bf A#1} #3 (#2)}
\def\jhep#1#2#3{J. High Energy Phys. {\bf #1} #3 (#2)}
\def\atmp#1#2#3{Adv. Theor. Math. Phys. {\bf #1} #3 (#2)}

\newcommand{\EQ}[1]{\begin{equation} #1 \end{equation}}
\newcommand{\AL}[1]{\begin{subequations}\begin{align} #1 \end{align}\end{subequations}}
\newcommand{\SP}[1]{\begin{equation}\begin{split} #1 \end{split}\end{equation}}
\newcommand{\ALAT}[2]{\begin{subequations}\begin{alignat}{#1} #2 \end{alignat}\end{subequations}}
\def\beq{\begin{equation}}
\def\eeq{\end{equation}}

\def\N{{\cal N}}
\def\sst{\scriptscriptstyle}
\def\thetabar{\bar\theta}
\def\Tr{{\rm Tr}}
\def\one{\mbox{1 \kern-.59em {\rm l}}}
 \def\Nh{\hat{N}}

\def\a{\alpha}      \def\da{{\dot\alpha}}
\def\b{\beta}       \def\db{{\dot\beta}}
\def\c{\gamma}  \def\G{\Gamma}  \def\cdt{\dot\gamma}
\def\d{\delta}  \def\D{\Delta}  \def\ddt{\dot\delta}
\def\e{\epsilon}        \def\vare{\varepsilon}
\def\f{\phi}    \def\F{\Phi}    \def\vvf{\f}
\def\h{\eta}
\def\k{\kappa}
\def\l{\lambda} \def\L{\Lambda}
\def\m{\mu} \def\n{\nu}
\def\o{\omega}
\def\p{\pi} \def\P{\Pi}
\def\r{\rho}
\def\s{\sigma}  \def\S{\Sigma}
\def\t{\tau}
\def\th{\theta} \def\Th{\Theta} \def\vth{\vartheta}
\def\X{\Xeta}
\def\z{\zeta}

\def\cA{{\cal A}} \def\cB{{\cal B}} \def\cC{{\cal C}}
\def\cD{{\cal D}} \def\cE{{\cal E}} \def\cF{{\cal F}}
\def\cG{{\cal G}} \def\cH{{\cal H}} \def\cI{{\cal I}}
\def\cJ{{\cal J}} \def\cK{{\cal K}} \def\cL{{\cal L}}
\def\cM{{\cal M}} \def\cN{{\cal N}} \def\cO{{\cal O}}
\def\cP{{\cal P}} \def\cQ{{\cal Q}} \def\cR{{\cal R}}
\def\cS{{\cal S}} \def\cT{{\cal T}} \def\cU{{\cal U}}
\def\cV{{\cal V}} \def\cW{{\cal W}} \def\cX{{\cal X}}
\def\cY{{\cal Y}} \def\cZ{{\cal Z}}

\def\ua{\underline{\alpha}}
\def\ub{\underline{\phantom{\alpha}}\!\!\!\beta}
\def\uc{\underline{\phantom{\alpha}}\!\!\!\gamma}
\def\um{\underline{\mu}}
\def\ud{\underline\delta}
\def\ue{\underline\epsilon}
\def\una{\underline a}\def\unA{\underline A}
\def\unb{\underline b}\def\unB{\underline B}
\def\unc{\underline c}\def\unC{\underline C}
\def\und{\underline d}\def\unD{\underline D}
\def\une{\underline e}\def\unE{\underline E}
\def\unf{\underline{\phantom{e}}\!\!\!\! f}\def\unF{\underline F}
\def\unm{\underline m}\def\unM{\underline M}
\def\unn{\underline n}\def\unN{\underline N}
\def\unp{\underline{\phantom{a}}\!\!\! p}\def\unP{\underline P}
\def\unq{\underline{\phantom{a}}\!\!\! q}
\def\unQ{\underline{\phantom{A}}\!\!\!\! Q}
\def\unH{\underline{H}}

\def\As {{A \hspace{-6.4pt} \slash}\;}
\def\bs {{b \hspace{-6.4pt} \slash}\;}
\def\Ds {{D \hspace{-6.4pt} \slash}\;}
\def\ds {{\del \hspace{-6.4pt} \slash}\;}
\def\ss {{\s \hspace{-6.4pt} \slash}\;}
\def\ks {{ k \hspace{-6.4pt} \slash}\;}
\def\ps {{p \hspace{-6.4pt} \slash}\;}
\def\pas {{{p_1} \hspace{-6.4pt} \slash}\;}
\def\pbs {{{p_2} \hspace{-6.4pt} \slash}\;}

\def\Fh{\hat{F}}
\def\Vh{\hat{V}}
\def\Xh{\hat{X}}
\def\ah{\hat{a}}
\def\xh{\hat{x}}
\def\yh{\hat{y}}
\def\ph{\hat{p}}
\def\xih{\hat{\xi}}

\def\psit{\tilde{\psi}}
\def\Psit{\tilde{\Psi}}
\def\tht{\tilde{\th}}
\def\lt{\tilde{\lambda}}
\def\At{\tilde{A}}
\def\Qt{\tilde{Q}}
\def\Rt{\tilde{R}}
\def\Nt{\tilde{N}}

\def\at{\tilde{a}}
\def\st{\tilde{s}}
\def\ft{\tilde{f}}
\def\pt{\tilde{p}}
\def\qt{\tilde{q}}
\def\vt{\tilde{v}}
\def\nt{\tilde{n}}

\def\delb{\bar{\partial}}
\def\bz{\bar{z}}
\def\bD{\bar{D}}
\def\bB{\bar{B}}

\def\bk{{\bf k}}
\def\bl{{\bf l}}
\def\bp{{\bf p}}
\def\bq{{\bf q}}
\def\br{{\bf r}}
\def\bx{{\bf x}}
\def\by{{\bf y}}
\def\bR{{\bf R}}
\def\bV{{\bf V}}

\def\d{\delta}\def\D{\Delta}\def\ddt{\dot\delta}

\def\pa{\partial} \def\del{\partial}
\def\xx{\times}
\def\uno{\mbox{1 \kern-.59em {\rm l}}}

\def\trp{^{\top}}
\def\inv{^{-1}}
\def\dag{{^{\dagger}}}
\def\pr{^{\prime}}
\def\lan{\langle}
\def\ran{\rangle}

\def\rar{\rightarrow}
\def\lar{\leftarrow}
\def\lrar{\leftrightarrow}

\newcommand{\0}{\,\!}      
\def\one{1\!\!1\,\,}
\def\im{\imath}
\def\jm{\jmath}

\newcommand{\tr}{\mbox{tr}}
\newcommand{\slsh}[1]{/ \!\!\!\! #1}

\def\vac{|0\rangle}
\def\lvac{\langle 0|}

\def\hlf{\frac{1}{2}}
\def\ove#1{\frac{1}{#1}}

\def\Box{\square}
\def\ZZ{\mathbb{Z}}
\def\CC#1{({\bf #1})}
\def\bcomment#1{}
\def\bfhat#1{{\bf \hat{#1}}}
\def\VEV#1{\left\langle #1\right\rangle}

\newcommand{\ex}[1]{{\rm e}^{#1}} \def\ii{{\rm i}}

\def\rr{{\rm r}} \def\rs{{\rm s}}\def\rv{{\rm v}}
\def\ri{{\rm i}}\def\rj{{\rm j}}

\newcommand{\lrbrk}[1]{\left(#1\right)}
\newcommand{\sfrac}[2]{{\textstyle\frac{#1}{#2}}}

\newcommand\equalhat{\mathrel{\stackon[1.5pt]{=}{\stretchto{%
    \scalerel*[\widthof{=}]{\wedge}{\rule{1ex}{3ex}}}{0.5ex}}}}

\font\mybb=msbm10 at 12pt
\def\bb#1{\hbox{\mybb#1}}

\font\myBB=msbm10 at 18pt
\def\BB#1{\hbox{\myBB#1}}

\newcommand{\tclr}{\textcolor}
\newcommand{\bpmat}{\begin{pmatrix}}
\newcommand{\epmat}{\end{pmatrix}}
\newcommand{\mrm}[1]{\mathrm{#1}}
\newcommand{\mrs}[1]{\scriptscriptstyle{\mathrm{#1}}}
\newcommand{\vct}[1]{\boldsymbol{#1}}
\newcommand{\hf}{\frac{1}{2}}
\newcommand{\x}{\times}
\newcommand{\pd}{\partial}
\newcommand{\dslash}{\displaystyle{\not}}
\newcommand{\ol}[1]{\overline{#1}}
\newcommand{\abs}[1]{\vert{#1}\vert}
\newcommand{\chiSqM}{\chi^2_{\mrm{min}}}
\newcommand{\chiSqMDof}{\chi^2_{\mrm{min}}/\mrm{d.o.f.}}
\newcommand{\om}{\omega}
\newcommand{\Lag}{\mathcal{L}}
\newcommand{\ord}{\mathcal{O}}
\newcommand{\eps}{\epsilon}
\newcommand{\beFrac}{\frac{1-\be}{1+\be}}
\newcommand{\beFracI}{\frac{1+\be}{1-\be}}
\newcommand{\amu}{a_{\mu}}
\newcommand{\damu}{\delta\amu}
\newcommand{\Damu}{\Delta\amu}
\newcommand{\amuUnit}{10^{-10}}
\newcommand{\mmu}{m_{\mu}}
\newcommand{\amuQED}{\amu^{\mrm{QED}}}
\newcommand{\amuEW}{\amu^{\mrm{EW}}}
\newcommand{\amuEWl}{\amu^{\mrm{EW,}\,1l}}
\newcommand{\amuEWll}{\amu^{\mrm{EW,}\,2l}}
\newcommand{\amuh}{\amu^{\mrm{had}}}
\newcommand{\amuhLO}{\amu^{\text{had, LOVP}}}
\newcommand{\amuhHO}{\amu^{\text{had, HOVP}}}
\newcommand{\amuhHOa}{\amu^{\text{had, HOVP(a)}}}
\newcommand{\amuhHOb}{\amu^{\text{had, HOVP(b)}}}
\newcommand{\amuhHOc}{\amu^{\text{had, HOVP(c)}}}
\newcommand{\amuhLbL}{\amu^{\text{had, LbL}}}
\newcommand{\ff}[3]{\mathcal{F}_{\pi^{0{#1}}\gamma^{#2}\gamma^{#3}}}
\newcommand{\alps}{\alpha_s}
\newcommand{\asmz}{\alpha_s(M_Z^2)}
\newcommand{\amz}{\alpha(M_Z^2)}
\newcommand{\aqmz}{\alpha_{\mrm{QED}}(M_Z^2)}
\newcommand{\delAlp}{\Delta\alpha}
\newcommand{\dAlpL}{\delAlp_{\mrm{lep}}}
\newcommand{\dAlpT}{\delAlp_{\mrm{top}}}
\newcommand{\dAlpH}{\delAlp_{\mrm{had}}}
\newcommand{\dAlpHF}{\dAlpH^{(5)}}
\newcommand{\dAlpHFmz}{\dAlpHF(M_Z^2)}
\newcommand{\tmin}{t_{\mrm{min}}}
\newcommand{\sTh}{s_{\mrm{th}}}
\newcommand{\eTh}{\sqrt{\sTh}}
\newcommand{\Ekmi}{E^{\,(k,m)}_i}
\newcommand{\Nkm}{N^{(k,m)}}
\newcommand{\Nkn}{N^{(k,n)}}
\newcommand{\Nexp}{N_{\mrm{exp}}}
\newcommand{\Nclu}{N_{\mrm{clu}}}
\newcommand{\Ntot}{N_{\mrm{tot}}}
\newcommand{\Rkmi}{R^{\,(k,m)}_i}
\newcommand{\Rknj}{R^{\,(k,n)}_j}
\newcommand{\dRkmi}{\mrm{d}\Rkmi}
\newcommand{\dRtkmi}{\mrm{d}\tilde{R}^{\,(k,m)}_i}
\newcommand{\BR}[2]{\mathcal{B}(#1\to #2)}
\newcommand{\decay}[2]{#1\to #2}
\newcommand{\UpsIVs}{\Upsilon(4S)}
\newcommand{\Gee}{\Gamma_{ee}}
\newcommand{\Gtot}{\Gamma_{\mrm{tot}}}
\newcommand{\ppC}{\pi^+\pi^-}
\newcommand{\ppN}{\pi^0\pi^0}
\newcommand{\pppC}{\pi^+\pi^-\pi^0}
\newcommand{\kkC}{K^+K^-}
\newcommand{\kskl}{K^0_S K^0_L}
\newcommand{\ksks}{K^0_S K^0_S}
\newcommand{\klkl}{K^0_L K^0_L}
\newcommand{\kskp}{K^0_S K^{\pm}\pi^{\mp}}
\newcommand{\eeMuMu}{e^+e^-\to\mu^+\mu^-}
\newcommand{\eeHadr}{e^+e^-\to\mrm{hadrons}}
\newcommand{\eeGhadr}{e^+e^-\to\gamma^*\to\mrm{hadrons}}
\newcommand{\tauNuHadr}{\tau\to\nu_{\tau}+\mrm{hadrons}}
\newcommand{\eeGPiPi}{e^+e^-\to\gamma^*\to\pi^+\pi^-}
\newcommand{\tauNuWNuPiPi}{\tau\to\nu_{\tau}W\to\nu_{\tau}\pi\pi^0}
\newcommand{\eeGIncl}{e^+e^-\to\gamma^*\to\mrm{all\,hadrons}}
\newcommand{\eeIncl}{e^+e^-\to\mrm{all\,hadrons}}
\newcommand{\eePiG}{e^+e^-\to\pi^0\gamma}
\newcommand{\eePiPi}{e^+e^-\to\pi^+\pi^-}
\newcommand{\eePiPiPi}{e^+e^-\to\pi^+\pi^-\pi^0}
\newcommand{\eeKK}{e^+e^-\to K^+K^-}
\newcommand{\ch}{\mrm{ch}}
\newcommand{\iso}{\mrm{iso}}
\newcommand{\noeta}{\text{no }\eta}
\newcommand{\kkr}{K\bar{K}\rho}
\newcommand{\kkp}{K\bar{K}\pi}
\newcommand{\kkpp}{K\bar{K}2\pi}
\newcommand{\kkppp}{K\bar{K}3\pi}
\newcommand{\isoAA}{(2\pi^+2\pi^-\pi^0)_{\mrm{no}\,\eta}}
\newcommand{\isoAB}{(\pi^+\pi^-3\pi^0)_{\mrm{no}\,\eta}}
\newcommand{\isoAC}{\omega(\to\mrm{npp})2\pi}
\newcommand{\isoACf}{\omega(\to\text{non-pure pionic states})2\pi}
\newcommand{\isoAD}{\eta\pi^+\pi^-}
\newcommand{\isoBA}{(2\pi^+2\pi^-2\pi^0)_{\mrm{no}\,\eta}}
\newcommand{\isoBB}{(\pi^+\pi^-4\pi^0)_{\mrm{no}\,\eta}}
\newcommand{\isoBC}{3\pi^+3\pi^-}
\newcommand{\isoBD}{\omega(\to\mrm{npp})3\pi}
\newcommand{\isoBDf}{\omega(\to\text{non-pure pionic state})3\pi}
\newcommand{\isoBE}{\eta\omega}
\newcommand{\isoEA}{\kkppp}
\newcommand{\isoEAa}{(K^+K^-\pi^+\pi^-\pi^0)_{\mrm{no}\,\eta}}
\newcommand{\isoEAb}{(K^0\bar{K}^0\pi^+\pi^-\pi^0)_{\mrm{no}\,\eta}}
\newcommand{\isoEB}{\omega(\to\mrm{npp})K\bar{K}}
\newcommand{\isoEBf}{\omega(\to\text{non-pure pionic states})K\bar{K}}
\newcommand{\isoEC}{\eta\phi}
\newcommand{\isoFA}{\eta2\pi^+2\pi^-}
\newcommand{\isoFB}{\eta\pi^+\pi^-2\pi^0}
\newcommand{\sigEEhadr}{\sigma(\eeHadr)}
\newcommand{\sigHad}{\sigma_{\mrm{had}}}
\newcommand{\sigHadB}{\sigHad^0}
\newcommand{\sigPt}{\sigma_{\mrm{pt}}}
\newcommand{\Rhad}{R_{\mrm{had}}}

\begin{document}

\title{\vspace{-3cm}The Muon $g-2$ experiment at Fermilab}

\author{\firstname{Alexander} \lastname{Keshavarzi}\inst{1}\fnsep\thanks{\email{aikeshav@olemiss.edu}} \\ {\small
        \firstname{on behalf of the} \lastname{Muon $g-2$ collaboration}\thanks{\texttt{http://muon-g-2.fnal.gov/collaboration.html}}} 
}

\institute{Department of Physics and Astronomy, The University of Mississippi, Mississippi 38677, U.S.}

\abstract{%
The current $\sim3.5\sigma$ discrepancy between the experimental measurement and theoretical prediction of the muon magnetic anomaly, $a_{\mu}$, stands as a potential indication of the existence of new physics. The Muon $g-2$ experiment at Fermilab is set to measure $a_{\mu}$ with a four-fold improvement in the uncertainty with respect to previous experiment, with an aim to determine whether the $g-2$ discrepancy is well established. The experiment recently completed its first physics run and a summer programme of essential upgrades, before continuing on with its experimental programme. The Run-1 data alone are expected to yield a statistical uncertainty of 350 ppb and the publication of the first result is expected in late-2019.
\vspace{-0.75cm}
}
\maketitle
\section{Introduction}\label{Sec:Introduction}
The muon magnetic anomaly, $a_{\mu} = (g-2)_{\mu}/2$, endures as an long-standing test of the Standard Model (SM), where the $\sim3.5\sigma$ (or higher) discrepancy between the experimental measurement $a_{\mu}^{\rm exp}$ and the SM prediction $a_{\mu}^{\rm SM}$ could be an indication of the existence of new physics beyond the SM. For $a_{\mu}^{\rm SM}$, work is ongoing to improve the estimates from all sectors of the SM, where in particular the hadronic contributions limit the total precision. Currently, the efforts of the {\em Muon $g-2$ Theory Initiative}~\cite{TGm2} and the groups involved within it show great progress and promise in improving the estimate of $a_{\mu}^{\rm SM}$ and its uncertainty. A recent precise re-evaluation of the dominating hadronic vacuum polarisation contributions to $a_{\mu}^{\rm SM}$ has resulted in $a_{\mu}^{\rm SM} = (11\ 659 \ 182.04 \pm 3.56) \times 10^{-10}$~\cite{Keshavarzi:2018mgv} (for alternative evaluations of $a_{\mu}^{\rm SM}$, see other analyses as part of~\cite{TGm2}).
Comparing this with the current experimental world average value of $a_{\mu}^{\rm exp} = (11\ 659 \ 209.1 \pm 5.4_{\rm stat} \pm 3.3_{\rm sys}) \times 10^{-10}$~\cite{PDG2016} results in a deviation between theory and experiment of $\Delta a_{\mu} = (27.06 \pm 7.26)\times 10^{-10}$, corresponding to a $3.7\sigma$ discrepancy~\cite{Keshavarzi:2018mgv}. The value for $a_{\mu}^{\rm exp}$ is entirely dominated by the measurements made at the Brookhaven National Laboratory (BNL)~\cite{Bennett:2002jb}, which achieved an overall precision of 540 parts-per-billion (ppb). However, efforts are currently underway to improve the experimental estimate at the Muon $g-2$ experiment at Fermilab (E989)~\cite{Grange:2015fou} and also at the proposed future J-PARC experiment~\cite{Mibe:2010zz}.

With the BNL measurement being statistics-limited, the goal of the E989 experiment is to measure $a_\mu$ with 20 times higher statistics, whilst also achieving a 2-3 times improvement in the systematic uncertainty~\cite{Grange:2015fou}. The target total uncertainty is 140ppb, which is a factor $\sim4$ improvement with respect to BNL. Assuming the hypothetical situation of the new experimental measurement at Fermilab yielding the same mean value for $a_{\mu}^{\rm exp}$ as the BNL measurement but achieving the projected improvement in its uncertainty, the comparison with $a_{\mu}^{\rm SM}$ would result in a $g-2$ discrepancy of $\sim7\sigma$~\cite{Keshavarzi:2018mgv}.

The E989 experiment completed its first physics run, Run-1, during March - July 2018. Immediately following this, a summer programme of essential upgrades was implemented, to ensure that the experiment would deliver on both its statistics and systematics goals for Run-2 and beyond. After briefly describing the experimental techniques employed by the Muon $g-2$ experiment at Fermilab to measure $a_\mu$, this article will review the status of the experiment after the Run-1 data-taking period and discuss the outlook for the future.

\section{Principles of the E989 measurement}

\subsection{Overview}

Following the same methodology as the previous experiment at BNL, the Muon $g-2$ experiment at FNAL injects longitudinally polarised muons into a storage ring with a magnetic dipole field $\lvert\vec{B}\rvert\sim 1.45$ T. In order to determine $a_\mu$, the experiment measures two frequencies: the frequency $\omega_a$ at which the muon spin (polarisation) turns relative to its momentum and the value of the magnetic field normalised to the Larmor frequency of a free proton, $\omega_p$. 

Assuming a perfect vertical magnetic field, with a muon on the ideal orbit, the anomalous precession frequency $\vec{\omega}_a$ is defined as the difference between the spin frequency $\vec{\omega}_S$ and the cyclotron frequency $\vec{\omega}_C$. In the absence of any other external fields,
\beq \label{eq:omega_aBasic}
\vec{\omega}_a = \vec{\omega}_S -\vec{\omega}_C = \ \biggr(\frac{g-2}{2}\biggr)\frac{e}{mc}\vec{B} = a_\mu\frac{e}{mc}\vec{B}\, ,
\eeq
as shown in Figure~\ref{fig:omega_a}. Note that should the gyromagnetic ratio $g=2$ exactly, then it would follow that $\vec{\omega}_S = \vec{\omega}_C$ such that the muon spin would precess with the same frequency as the orbital frequency, resulting in $a_\mu = 0$. 

Determining $a_\mu$ according to equation~\eqref{eq:omega_aBasic} requires evaluating the average magnetic field experienced by the stored muons. Therefore, the $\vec{B}$-field must be convoluted with the muon distribution by integrating all measured values of $\omega_p$ over the muon storage region. Once both $\omega_a$ and $\omega_p$ have been correctly extracted, the muon magnetic anomaly is calculated via~\cite{Grange:2015fou}
\beq \label{eq:expamufinal}
a_\mu = \frac{g_e}{2}\frac{m_\mu}{m_e}\frac{\mu_p}{\mu_e}\frac{\omega_a}{\omega_p} \, ,
\eeq
where $g_e$ is the gyromagnetic ratio of the electron, $m_\mu/m_e$ is ratio of the muon and electron masses and $\mu_p/\mu_e$ is the measured ratio of the proton and electron magnetic moments.


\subsection{Producing and storing the muon beam}

The accelerator complex at Fermilab was chosen due to its capability of producing a high-purity, intense muon beam, which provides an outstanding statistics advantages compared to the BNL experiment. An 8 GeV proton beam is fired at a target to produce pions. These pions are then injected into a delivery ring, in which the pions decay into muons via \allowbreak $\pi^+~\rightarrow~\mu^+~\nu_{\mu}$. A long decay channel for the pion decay into muons is provided by $\sim 4$ orbits of the delivery ring. During this time, any remaining protons and the resulting muons are separated by enough orbital distance that the protons can be safely discarded from the beam. The result is a $\sim96\%$ longitudinally-polarised $\mu^+$ beam consisting of $8\times10^6$ muons/sec with an energy of 3.094 GeV/c and a $\pi^+$ content of $<10^{-5}\%$. Not only does this allow for 21 times more decay positrons to be detected than at BNL, but the minimal proton and pion contamination also reduces the hadronic flash recorded by the detectors at beam injection by a factor of 20 compared to the previous experiment.

The BNL storage ring has been recommissioned for the Fermilab experiment. In 2013, it was transported 3,200 miles on a high-profile 35-day journey that received great interest from the wider physics community and the general public. To safely the deliver the muons to the experiment from the upstream beam-line, the beam is injected into the storage ring through a superconducting inflector magnet, which locally cancels the $\sim1.45$ T storage $\vec{B}$-field to allow the muons to enter without being deflected. 

\begin{figure}[!t]
\centering
\subfloat[]{%
\includegraphics[width= 0.305\textwidth]{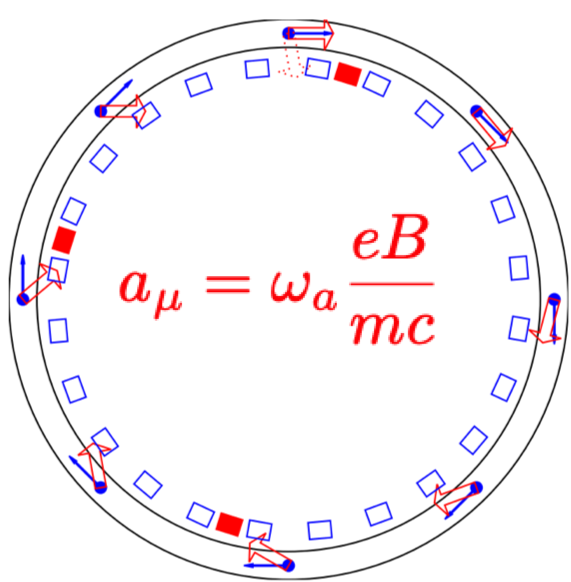}\label{fig:omega_a}}\hspace{1.5cm}
\subfloat[]{ %
\includegraphics[width= 0.325\textwidth]{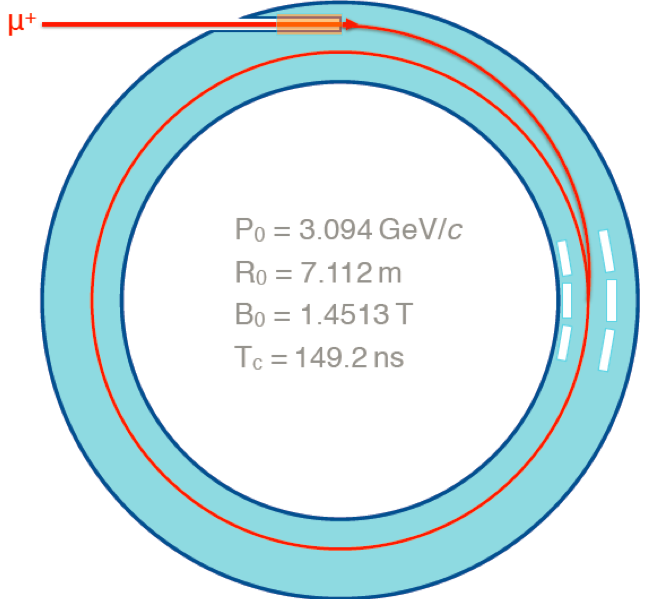}\label{fig:kick}}\hfill
\caption{(a) The spin precession of muons utilised in order to measure $a_{\mu}$. The blue arrows show the muon momentum, whilst the red arrows indicate the muon spin (and corresponding decay positron direction). (b) The injection and storage of muons, where the 3 kicker plates are shown in white at 90 degrees from the injection point. Here, ${\rm P_0}$ is the "magic" muon momentum, ${\rm R_0}$ is the orbit radius, ${\rm B_0}$ is the field magnitude of the storage ring magnet and ${\rm T_c}$ is the cyclotron period.}
\vspace{-0.75cm}
\end{figure}
As can be seen from Figure~\ref{fig:kick}, the muon beam is injected onto an orbit that is displaced by 11 mrads radially outward from the ideal orbit of the storage ring. In order to direct the beam back onto the correct trajectory requires a 'kick'. The requirements of each of the three kickers are to produce a $\sim300$ Gauss field over 4 metres for 120 ns at 100 Hz as the muon beam pulse traverses through the domain of the kicker region. To provide vertical focusing, electrostatic quadrupoles are used to confine the muon beam in the storage region. The application of an electric field introduces a new term to equation \eqref{eq:omega_aBasic}, where relativistic particles feel a motional magnetic field proportional to $\vec{\beta}\times\vec{E}$. Here, $\vec{\beta}$ denotes the muon velocity and $\vec{E}$ is the electric field. In full, the equation for $\vec{\omega}_a$ is
\beq\label{omega_a} 
\vec{\omega}_a = \ \frac{e}{mc}\Biggr[a_{\mu}\vec{B} \ - \ \Biggr(a_\mu-\frac{1}{\gamma^2-1}\Biggr)\ \vec{\beta}\times\vec{E} \ - \ a_\mu \ \Biggr(\frac{\gamma}{\gamma+1}\Biggr)\ (\vec{\beta}\cdot\vec{B}) \ \vec{\beta}\ \Biggr]\,,
\eeq
where the third term additionally accounts for those muons whose motion is not perpendicular to the magnetic field. To first order, the second term in equation~\eqref{omega_a} vanishes for the choice of muons at the "magic" momentum of 3.094 GeV/c~\cite{Bailey:1978mn} and the majority of otff-momentum muons are removed using collimators. However, there persists a small momentum spread of remaining muons away from the magic momentum and some also experience a small amount of vertical pitching, corresponding to the second and third terms of equation~\eqref{omega_a} respectively. The magnitudes of these effects are determined via data analysis, allowing for well-known, sub-parts-per-million (ppm) systematic corrections and corresponding uncertainties to be applied to the measured $\omega_a$.


\subsection{Measuring the anomalous precession frequency, $\omega_a$}\label{sec:omega_a}

In the storage ring, muons decay via the parity-violating weak process $\mu^+ \rightarrow e^+\bar{\nu}_\mu\nu_e$ into positrons. The higher-energy positrons are preferentially emitted along the direction of the muon spin, such that the detection of the arrival time and energy of the decay positrons above an appropriate energy cut can be used to infer the spin direction and extract $\omega_a$. The primary detectors for this purpose are 24 calorimeters placed at equidistant positions on the inner radius of the storage ring, as shown by the blue rectangles in Figure~\ref{fig:omega_a}. These calorimeters record the oscillation in the number of detected positrons over time due to the spin precession. An example of this for reconstructed Run-1 data is shown in Figure~\ref{fig:60hour} in Section~\ref{sec:Run1}.

In its simplest form, the analysis technique for the extraction of $\omega_a$ follows by fitting the data to the a five-parameter function of the form
\beq\label{eq:5par}
f(t) = Ne^{-t/\tau}[1+A\cos(\omega_at + \phi)]\, .
\eeq
where $N$ is the overall normalisation, $\tau$ is the boosted muon lifetime, $A$ is the overall muon asymmetry and $\phi$ is the initial phase. There are several subtle variations and/or analyser choices possible when attempting this fitting procedure, resulting in many different analysis groups analysing the same data via alternate methods, each providing a consistency check of the others. 

Due to the storage environment, the five-parameter function in equation~\eqref{eq:5par} is not sufficient in describing all the relevant beam dynamics that contribute to the oscillation in the data rate that is detected. For example, the injection of the muon beam at a radial offset and an imperfect kick lead to coherent betatron oscillations (CBO) of the beam in the radial direction. The calorimeters are sensitive to this and other similar systematic effects. Therefore, extra terms are necessarily added to equation~\eqref{eq:5par} to describe the additional beam dynamics and corresponding systematic uncertainties are evaluated. Other detector systems are also utilised that are invaluable in this regard. Two straw tracker stations, for example, measure the spatial profile of the stored muons, which help to determine the values of the two corrections (the $E$-field and pitch corrections) in equation~\eqref{omega_a}, monitor the beam oscillations and measure the muon distribution to convolute with the magnetic field measurement. Additionally, a state-of-the-art laser calibration system has been developed that monitors and provides calorimeter gain stability to sub-per-mil accuracy~\cite{Anastasi:2016luh}.

\begin{figure}[!t]
\centering
\subfloat[$\omega_a$ systematic uncertainty budget.]{%
\includegraphics[width= 0.49\textwidth]{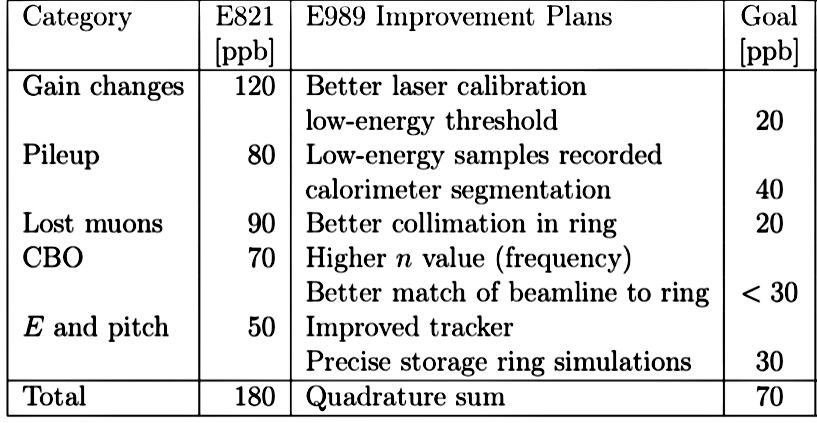}\label{fig:omega_aSys}}
\subfloat[$\omega_p$ systematic uncertainty budget.]{ %
\includegraphics[width= 0.5\textwidth]{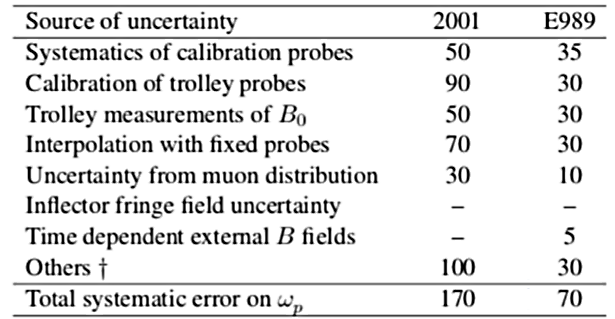}\label{fig:omega_pSys}}\hfill
\caption{The design systemic uncertainty goals of the measurements of $\omega_a$ and $\omega_p$ compared the final systematic uncertainty estimate from the BNL experiment~\cite{Grange:2015fou}. All values are given in ppb.}
\vspace{-0.75cm}
\end{figure}

The uncertainty budgets of the $\omega_a$ measurement over the course of the entirety of the Muon $g-2$ experiment are expected to be 100 ppb statistical uncertainty and 70 ppb systematic uncertainty. The final systematic uncertainty goals for the measurement of $\omega_a$ are displayed in Figure~\ref{fig:omega_aSys}.


\subsection{Measuring the magnetic field in terms of the proton NMR frequency, $\omega_p$}\label{sec:omega_p}

\begin{figure}[!t]
\centering
\includegraphics[width= 0.7\textwidth]{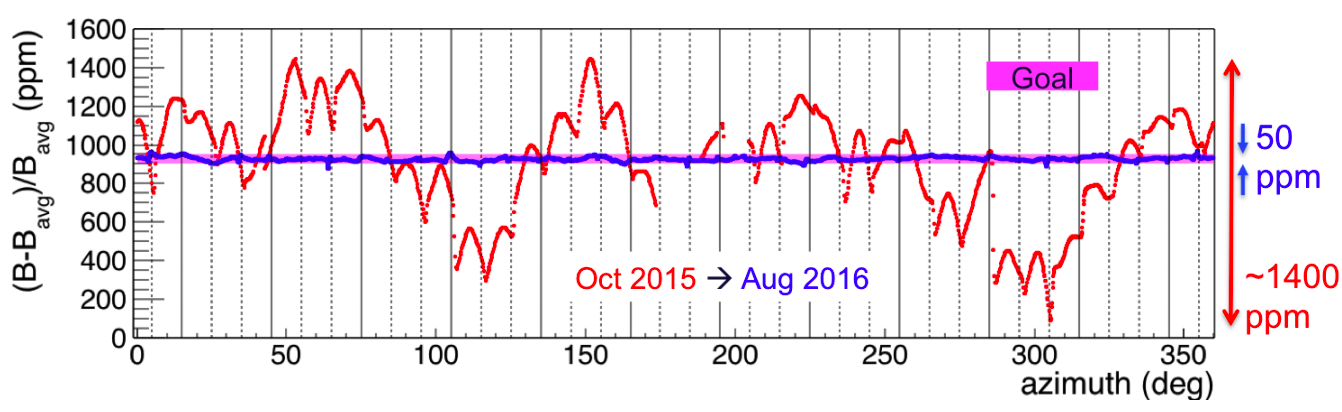}
\caption{Measurements of the uniformity of the magnetic dipole field before (red) and after (blue) shimming.}\label{fig:shimming}
\vspace{-0.75cm}
\end{figure}

The uniformity of the field of the storage ring magnet was achieved over an almost year-long magnetic shimming process and is continuously maintained, with a field homogeneity of $\pm 40$ ppm in any remaining variations. Measurements of the dipole field uniformity before and after the shimming programme are displayed in Figure~\ref{fig:shimming}. The $B$-field is monitored continuously using 400 inner-chamber, fixed NMR probes located above and below the storage region and is periodically measured by a trolley that traverses the storage region itself, mapping the magnetic field experienced locally by the muons using 17 additional NMR probes. For the contribution to $a_\mu$, the absolute field value is determined from the measured $\omega_p$ using absolute field calibration probes and is convoluted with the muon distribution as measured by the trackers. The systematic uncertainties goals for the measurement of $\omega_p$ are displayed in Figure~\ref{fig:omega_pSys}.


\subsection{Analysis and hardware blinding}

The Muon $g-2$ experiment is both hardware and software blinded. The 40MHz that clock drives the calorimeter digitizers, straw tracker and NMR digitisers has had an offset in the range $\pm25$ ppm applied to externally blind the time of the measured $\omega_a$. Software blinding occurs at the analysis stage for both the $\omega_a$ and $\omega_p$ frequencies. In the $\omega_a$ analysis, analysers actually fit for the quantity $R$ given in the expression
\beq
\omega_a = 2\pi\cdot 0.2291 {\rm MHz} \cdot [1-(R-\Delta R) \times 10^{-6}]\, ,
\eeq
where $\Delta R$ is a frequency offset that is unique to each analyser. In late stages of the analysis, before a final unblinding, relative unblinding exercises of these individual offsets to a common offset are performed for all analyses and datasets, allowing for internal comparison and consistency checks without the risk of conscious or unconscious bias.


\section{Current experimental status following Run-1} \label{sec:Run1}

\begin{figure}[!t]
\centering
\subfloat[Total number of positrons collected during Run-1 before quality cuts.]{%
\includegraphics[width= 0.4625\textwidth]{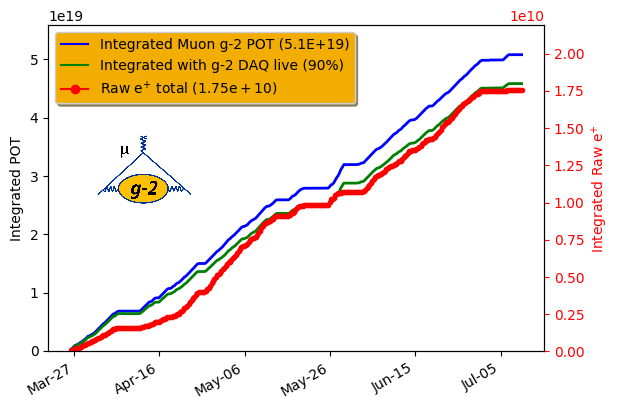}\label{fig:CTAG-Run1}}\hspace{0.5cm}
\subfloat[The '60 hour' data set.]{ %
\includegraphics[width= 0.44\textwidth]{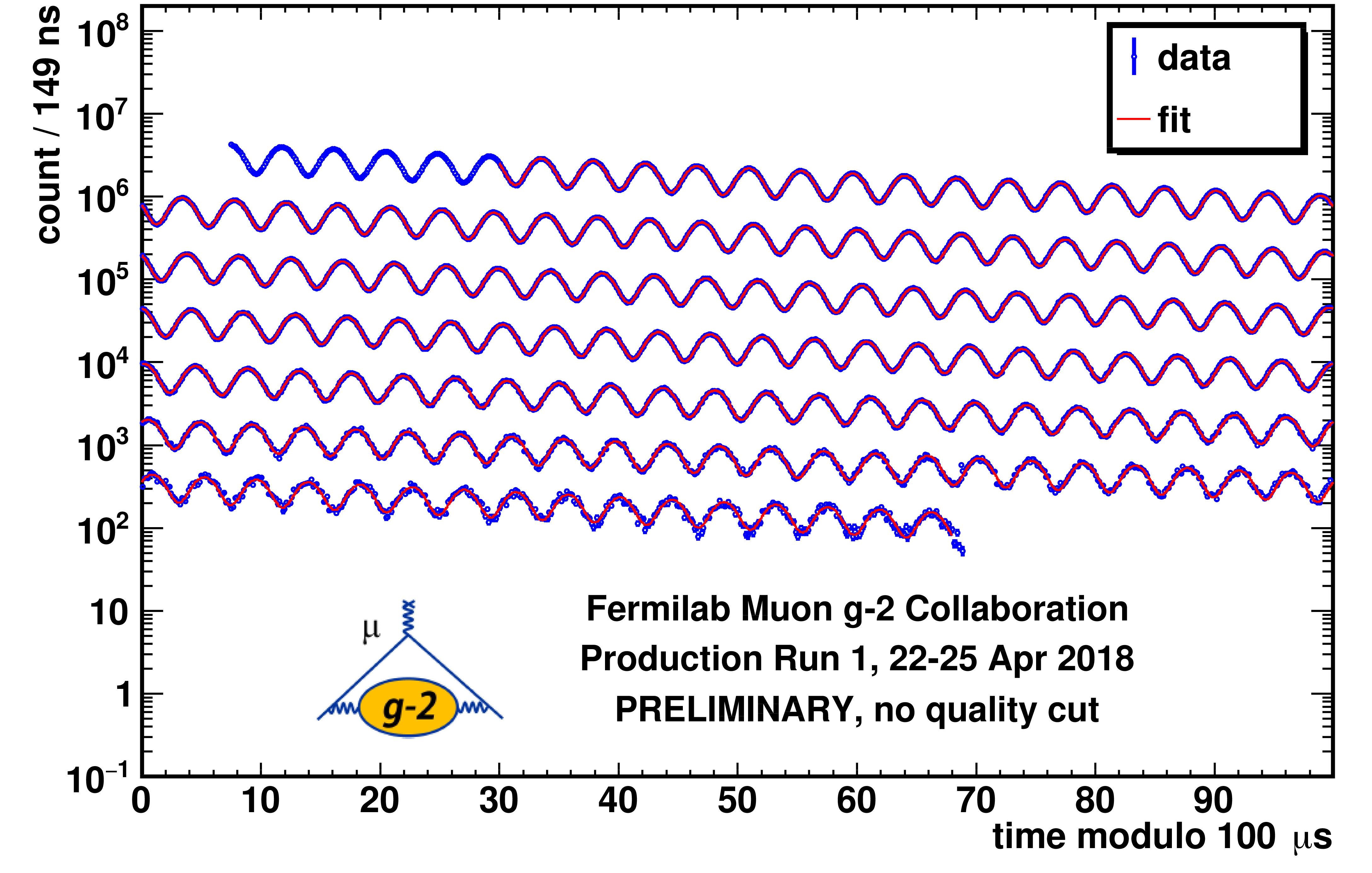}\label{fig:60hour}}\hfill
\caption{Data collected during Run-1 of the Muon $g-2$ experiment.}
\vspace{-0.75cm}
\end{figure}

During Run-1, the Muon $g-2$ experiment recorded 17.5 billion positrons, almost twice the number recorded from both the combined $\mu^+$ and $\mu^-$ runs of the BNL experiment. The analysable data after data quality cuts (DQC) is $1.38$ times that of the BNL dataset, with a projected statistical uncertainty of 350 ppb. Figure~\ref{fig:CTAG-Run1} shows the accumulation of data over the course of Run-1. Figure~\ref{fig:60hour} shows the fit-to-data of 0.95 billion positrons accumulated over 60 hours in mid-April 2018, corresponding to a statistical uncertainty of 1.3 ppm. The publication of the result from all data from Run-1 is expected in late-2019.

Studies of the performance during Run-1 showed that the muon storage was roughly half that of the original design specifications. This was, in-part, due to an underperformance of the kicker system, which was found to have a 30\% deficit in required kick strength. Therefore, the summer shutdown that directly followed the end of Run-1 involved a substantial kicker system upgrade. Additional upgrades were performed to the electrostatic quadrupoles and the accelerator complex. The installation of a new open-ended inflector magnet has been planned for the end of Run-2 in the summer of 2019, which is projected to incur a further $\sim 40\% $ increase in the number of stored muons. Currently, the Muon $g-2$ experiment is due to continue running for a further two years to obtain the full $20\times$ BNL statistics dataset.




\section{Conclusions and outlook}

The Muon $g-2$ experiment at Fermilab is undertaking the formidable task of measuring $a_\mu$ with a target total uncertainty of 140ppb. The comparison of this measurement with the impressive SM prediction $a_\mu^{\rm SM}$ will determine whether the current discrepancy between theory and experiment of $\sim 3.5\sigma$ is well established. In in order to evaluate $a_\mu$, the E989 experiment follows the principles of the BNL experiment to measure the muon anomalous precession frequency $\omega_a$ and the magnetic field in terms of $\omega_p$. The first physics run and a summer programme of essential upgrades are already complete. During Run-1 alone, the experiment collected twice the number of positrons collected at the BNL experiment, which after DQC should yield a statistical uncertainty of 350 ppb. The first publication of the result from Run-1 is expected in late-2019.

\section*{Acknowledgements}

{
The author would like to thank the organisers of {\em International Workshop on $e^+e^-$ collisions from Phi to Psi (PhiPsi19)} for a very productive and enjoyable workshop. This manuscript has been authored by an employee of The University of Mississippi, supported in-part by the U.S. Department of Energy Office of Science, Office of High Energy Physics, award DE-SC0012391. This document was prepared by the Muon $g-2$ collaboration using the resources of the Fermi National Accelerator Laboratory (Fermilab), a U.S. Department of Energy, Office of Science, HEP User Facility. Fermilab is managed by Fermi Research Alliance, LLC (FRA), acting under Contract No. DE-AC02-07CH11359.

}

%
%
%
{
}

\end{document}